\documentclass[twocolumn,showpacs,preprintnumbers,amsmath,amssymb]{revtex4}
\usepackage{graphicx}% Include figure files
\usepackage{dcolumn}% Align table columns on decimal point
\usepackage{bm}% bold math

\begin{document}

%\draft
\preprint{ }
\title{
Accurate and concise atomic CI via generalization of analytic
Laguerre type orbitals and examples of \emph{ab-initio} error
estimation for excited states}
\author{
Z. Xiong$^{1,2}$ and N.C. Bacalis$^{2}$}\email{nbacalis@eie.gr}
\affiliation{
$^{1}$Engineering Science Department, University of
Patras, GR-26500 Patras, Greece. \\
$^{2}$Theoretical and Physical Chemistry Institute, National
Hellenic Research Foundation, Vasileos Constantinou 48, GR - 116
35 ATHENS, Greece.
}

\date{\today}

\begin{abstract}
We propose simple analytic, non-orthogonal but selectively
orthogonalizable, generalized Laguerre type atomic orbitals,
providing clear physical interpretation and \emph{near equivalent
accuracy with numerical multi-configuration self-consistent
field}, to atomic configuration interaction calculations. By
analyzing the general Eckart theorem we use their simple
interpretation, via a thorough investigation in orbital space, to
estimate, for the first time (the exact value being, or
considered, unknown), an \emph{ab-initio energy uncertainty}, i.e.
proximity to the exact energy, for several excited atomic states
known to have the danger to suffer from variational collapse.
\end{abstract}

\maketitle
% body of paper here
\section{Outline}
The purpose of this paper is threefold. (i) First is to show that,
in variational \emph{ab-initio} atomic configuration interaction
(CI) calculations (for the ground or excited states), \emph{by
varying the extent and the node positions of  the (analytic)
orbital radial functions, it is possible to achieve nearly
numerical multi-configuration self-consistent field (NMCSCF)
accuracy}. (This means that the resulting analytic orbitals,
similar to NMCSCF orbitals, are few, concise, clearly
interpretable and with rich physical content). (ii) The second
purpose is to analyze and clarify an extension of the Eckart
theorem for excited states (c.f. Appendix). (iii) The third is to
demonstrate that, (with the help of these, at least, orbitals),
within the general Eckart theorem, it is possible to obtain an
\emph{ab-initio} \emph{estimate of the proximity to the
(supposedly unknown) exact energy}, at least in some special cases
of CI expansions. Consequently it is possible, in these cases,
that other outcomes (outside of the uncertainty error) be \emph{ab
initio} rejected without the need of external information.

\section{Introduction}

(i) Among various configuration interaction (CI) methods for the
calculation of the electronic structure of atoms, based on the
variational principle, the NMCSCF is very efficient and accurate
because it describes the electronic state with few simple
orbitals, rich in physical meaning, whereas other methods are
based on large basis sets which complicate the description. It
would be interesting to invent \emph{analytic} orbitals similar to
NMCSCF, thus describing the state in an equally simple and concise
way, with comparable accuracy. We have invented such analytic
semiorthogonal basis functions, which very satisfactorily
approximate the NMCSCF orbitals: The central idea is to adopt the
usual Laguerre orbitals and start moving variationally the nodes
and the extent of their radial functions, until minimization of
the energy. Eventually the resulting orbitals are similar and of
comparable accuracy with NMCSCF. Nevertheless, since NMCSCF is
numerical, it leads to an energy minimum, indifferently global or
local, usually the widest (which is sometimes misinterpreted
\cite{[scffail_b],[scffail_m]}) and other energy minima cannot be
easily located, which may describe the state in a simpler way,
i.e., with a smaller contribution of the higher order terms in the
CI expansion. For this reason, by looking at only the widest
minimum, the quality of the approximation to the exact
eigenfunction cannot be \emph{ab-initio} estimated. However, since
our orbitals are analytic, all (finite in number) energy minima
can be located, at least in principle, and because of the
simplicity, and of the immediate recognizability of the physical
content of the orbitals, the most representative description of
the CI expansion can be chosen, therefore a measure of the quality
of our approximation to the exact solution can be \emph{ab-initio}
estimated (without using other external information) via the
correction introduced by the general Eckart theorem (c.f.
Appendix). The analyticity of the orbitals allows also the
flexibility to have orthogonal occupied orbitals and
non-orthogonal some virtual correlation orbitals, thus
accelerating the convergence of the CI expansion.

(ii) On the other hand, for some excited states, like He 1s2s
$^1S$, Mg 3s4s $^1S$ etc \cite{[18]}, the variational calculation
may collapse to a lower lying than the exact state, which is
wrong, but \emph{allowed} by the general Eckart theorem for
excited states (c.f. Appendix). It would be interesting, if
possible, to invent an \emph{ab-intio} (without other external
information) way to reject such wrong outcomes and to
variationally bracket the unknown exact energy level within some
known digits of certainty. To this end we propose a method,
feasible with the presented orbitals, valid under certain
conditions, and demonstrate it in several cases with two or three
electrons. We also discuss the extent of its feasibility.

\section{Part I. Atomic CI via generalization of Laguerre type orbitals}

We propose a generalization of Laguerre type orbitals to the form
$\langle \textbf{r}|n,l,m\rangle$ $=$ $A_{n,l,m}$ $L_{n,l} (\{ g\}
,r,z_{n,l} ,b_{n,l} ,q_{n,l} )$ $Y_{l,m} (\theta ,\phi )$, where
$A_{n,l,m}$ is a normalization constant and $Y_{l,m} (\theta ,\phi
)$ are spherical harmonics. The generalized Laguerre type
functions (GLTOs) are (in a.u.)
\begin{widetext}
\begin{eqnarray}
%\begin{array}{l}
L_{n,l} (\{ g\} ,r,z_{n,l} ,b_{n,l} ,q_{n,l} )= \sum\limits_{k{\rm
= 0}}^{n - l - 1} {{\rm c}_k } (n,l,z_{n,l} )g_k (n,l,\{ z_{i,l}
,b_{i,l} ,q_{i,l} \}_{i = 1}^n )r^{l + k}\exp (-z_{n,l}r/n)\nonumber\\
 + b_{n,l} \exp (-q_{n,l}z_{n,l}r/n)\delta _{l,0}
%\end{array}
\label{orb}
\end{eqnarray}
\end{widetext}
where $g_{n - l -1}(n,l,\{z_{i,l},b_{i,l},q_{i,l}\}_{i=1}^n)$ $=
1$ (about the rest of the $g_k$ factors we extensively discuss
below) and $c_{k}$ are the usual associated Laguerre polynomial
coefficients. The parameters $z_{n,l}$, $b_{n,l}$, $q_{n,l}$ are
determined from the (non-linear variational) minimization
\cite{[10]} of the desired root of the secular equation (see below
equation (\ref{det})). The $z_{n,l}$ parameters are effective
nuclear charges and determine the radial extent of the orbitals.
Since $z_{n,l}$ differ from orbital to orbital, these orbitals
are, in general, \emph{non-orthogonal}. The addition of the last
term of equation (\ref{orb}) just modifies the radial part of
$s$-orbitals, since 1s cannot be modified by any $g_k$ factor.

Then, a normalized CI wave function is formed out of Slater
determinants (composed of the proposed orbitals), whose node
positions and radial extent are optimized variationally through
non-linear multidimensional minimization of the total energy. We
present a selective intrinsic orthogonalization formalism to any
lower $n$,$l$ orbital of either the ground, or a desired excited
state, thus \emph{preserving} the orbital characteristics.
\emph{The rest of the orbitals remain non-orthogonal} (e.g., see
table \ref{tab:orb} below). We first find (and use) a main wave
function in the dominant part of the active space (called
`\emph{main}'), well representing the state under consideration
[e.g., for He, in the active space of $2s$, $2p$, $3s$, $3p$, the
four $^1 P ^o$ roots have the following `\emph{main}' wave
functions: ($2s2p$), ($2s3p\pm3s2p$) and ($3s3p$)], and then we
add angular and radial correlation \cite{[2]}, simulating cusp
conditions. The method is tested against several known cases.

Thus, given the atom with nuclear charge $Z_{nuc}$, and $N$
electrons, with space and spin coordinates
$\textbf{r}_{1}s_{1},...,\textbf{r}_{N}s_{N}$, as well as the
symmetry and the electron occupancy, the desired $N-$electron
normalized wave function, consisting of $N_c$ (predetermined)
configurations, out of $N_d$ Slater determinants, is
\begin{eqnarray}
\begin{array}{l}
\Psi ({\bf r}_1 s_1 ,...,{\bf r}_N s_N ) = \sum\limits_{p =
1}^{N_c} {c_p \Gamma _p }
({\bf r}_1 s_1 ,...,{\bf r}_N s_N ); \\
\left| \Psi \right|^2 = 1;\Gamma _p =
\sum\limits_{i = 1}^{N_d} {f_{ip} } D_i \\
\end{array}\label{wfn}
\end{eqnarray}
where the linear parameters $c_{p}$ are determined from a desired
(usually the lowest) root of the secular equation
\begin{eqnarray}
\det \left[\left( \sum\limits_{a,b = 1}^{N_d} {f_{ap} f_{bq} }
\left\langle {D_a } \right|H - E\left| {D_b } \right\rangle
\right)_{p,q} \right]_{N_c\times N_c} = 0 \label{det}
\end{eqnarray}
which is solved by the strategy of p. 455 of ``Numerical Recipes''
\cite{[6]}. Here $E$ is the total energy, the Hamiltonian matrix
elements are calculated by the method of p. 66 of McWeeny
\cite{[7]}, where
\begin{eqnarray}
H&&= - \frac{1}{2}\sum\limits_{i = 1}^{N} {(\nabla _i^2 +
\frac{{Z_{nuc} }}{{\left| {{\bf r}_i } \right|}}} )
 + \sum\limits_{i > j}^{N} {\frac{1}{{\left|
 {{\bf r}_i - {\bf r}_j } \right|}}} \nonumber \\
&&~\equiv \sum\limits_{i = 1}^{N}h(i) + \sum\limits_{i >
j}^{N}g(i,j) ; \label{ham}
\end{eqnarray}
the $D_{a}$ are all ($N_d$) (consistent with the desired
electronic state) Slater determinants, formed out of $N_{orb}$
(predetermined), to be optimized, spinorbitals $a_{i}$, and
$f_{ap}$ are all ($N_d \times N_c$) consistent corresponding
coefficients, which we determine by implementing the ideas of
Schaefer and Harris's method \cite{[8]}. The angular (and spin)
part of the matrix elements in equation (\ref{det}) we treat
according to the method of chapter 6 of Tinkham \cite{[9]}.

The adaptability of our orbitals to almost NMCSCF accuracy is due
to the $g_k$-factors, which move, during the minimization process,
the orbital nodes appropriately, by intrinsic orthogonalization
among \emph{desired} orbitals of the same \cite{[11]} or of a
different \cite{[5]} state (an advantage of this method), by
directly solving $\langle n_i ,l,m |n_j ,l,m\rangle$ $=$ $\delta
_{i,j}$, $(i, j =1,..., N_{orb})$. For example, for the $1s$ and
$2s$ orbitals, equation $ \langle 1s | 2s \rangle = 0$ yields
\begin{widetext}
\begin{eqnarray}
g_0 (2,s,\{ z_{1s} ,z_{2s} \} ) = {\frac{{\frac{6\,z_{2s}}
{{{\left(2\,z_{1s}+z_{2s}\right)}^4}}}+{\frac{6\,b_{1s}\,
z_{2s}}{{{\left(2\, q_{1s}\, z_{1s}+z_{2s}\right)}^4}}}-
{\frac{b_{2s}}{{{\left(2\,z_{1s}+q_{2s}\, z_{2s}\right)}^3 }}}-
{\frac{b_{1s}\,b_{2s}}{{{\left(2\, q_{1s}\, z_{1s}+q_{2s}\, z_{2s}
\right)}^3 }}}}{{\frac{2}{{{\left(2\,z_{1s}+z_{2s}\right)}^3}}}+
{\frac{2\,b_{1s}}{{{\left(2\,q_{1s}\, z_{1s}+z_{2s}\right)}^3}
}}}}\label{g2s}
\end{eqnarray}
\end{widetext}
and it is straightforward to derive the $g_k$-factors for
$n=2,3,4,...$, $l=0,1,2,...,n-1$ and $k=1,...,n-l-2$
\cite{[1],[12]}.

Thus these orbitals, after orthogonalization, \emph{are not linear
combinations of each other}, as in orther orthogonalization
schemes, but \emph{maintain (c.f. equation (\ref{orb}) and figure
(\ref{f1}) below) a clear physical interpretation for all} $l = s,
p, d,...$, enabling one, to choose reasonable (and to reject
unreasonable) outcomes even by inspection.

Since the CI expansion may still contain non-orthogonal orbitals,
we use the general non-orthogonal formalism of p. 66 of McWeeny
\cite{[7]},

\begin{eqnarray}
& \langle D_a|\sum\limits_{i = 1}^{N}h(i)|D_b\rangle = \nonumber\\
& (D_{aa} D_{bb})^{-1/2}\sum\limits_{i,j = 1}^{N}\langle a_{i}|h
|b_{j}\rangle D_{ab}(a_{i}b_{j}) \label{m11}
\end{eqnarray}
where $D_{ab} = det|\langle a_1|b_1\rangle \langle a_2|b_2\rangle
... \langle a_n|b_n\rangle|$, $D_{ab}(a_{i}b_{j})$ denotes the
cofactor of the element $\langle a_{i}|b_{j}\rangle$ in the
determinant $D_{ab}$, and $D_{aa},D_{bb}$ are similar
normalization factors; $a_{i}$, $b_{j}$ are the spinorbitals.
Also,

\begin{eqnarray}
& \langle D_a|\sum\limits_{i>j}^{N}g(i,j)|D_b\rangle = \nonumber\\
&(D_{aa}
D_{bb})^{-1/2}\sum\limits_{i>t}^{N}\sum\limits_{j>l}^{N}\langle
a_{i}a_{t}|g |b_{j}b_{l}\rangle D_{ab}(a_{i}a_{t}b_{j}b_{l})
\label{m12}
\end{eqnarray}
where $D_{ab}(a_{i}a_{t}b_{j}b_{l})$ is the cofactor of $D_{ab}$
defined by deleting the rows and columns containing $\langle
a_{i}|b_{j}\rangle$ and $\langle a_{k}|b_{l}\rangle$ and attaching
a factor $(-1)^{i+j+t+l}$ to the resultant minor. In principle,
equations (\ref{m11}-\ref{m12}) can readily deal with Slater
determinants for any large atom without leading to extra
complexity, so that one need not adopt ``limited
non-orthogonality" in order to avoid complications.

We improve the N-electron wave function by incorporating
\emph{radial} and \emph{angular} correlation %i.e. we enrich it
%with more \emph{nodes} and more \emph{lobes}, respectively,
so as to simulate the cusp conditions, either via
orthogonalization to desired lower-$n$ orbitals, or via free
non-orthogonality.

The contraction with the $l=0$ Slater type orbital (the last term
of equation (\ref{orb})), especially useful when there are outer
electrons \emph{repelling the inner toward the nucleus}, provides,
in full CI, about 75\% of the energy correction obtained if we
freely doubled the (uncontracted) orbitals, while it substantially
reduces the CI size; symbolically, $(E_{N}^{c}-E_{N}^{u})$ $\sim$
75\% $(E_{2N}^{u}-E_{N}^{u})$. Thus, for $Be$ $1s^{2}2s^{2}$
$^{1}S$, with 2 orbitals $1s$, $2s$ (1 configuration) we have (in
$a.u.$) $E_{2}^{u}$ $= -14.5300$, $E_{2}^{c}$ $ = -14.5723$, while
with 4 orbitals $1s, 1s', 2s, 2s'$ (20 configurations), $E_{4}^{u}
=$ $-14.5893 $; i.e., the contraction provides 71\% of the
corresponding (uncontracted) CI correction. Similarly, for the
$H^{-}$ $1s^{2}$ $^{1}S$, with 4 orbitals of $1s$ type with 10
configurations we obtain $E_{4}^{u}= -0.51438$, $E_{4}^{c}$ $ =
-0.51445 $, while 8 uncontracted $1s$ type orbitals with 36
configurations give $E_{8}^{u} =$ $ -0.51448$, i.e., the
contraction provides 74\% of the corresponding free CI correction.

In addition, we performed several further tests:

1. For the $He$ ground state $1s^{2}$ $^{1}S$, if we use the
uncontracted correlation orbitals of table (\ref{tab:orb}) up to
$4f'$, we obtain (in $a.u.$) $E = -2.903104$, which is comparable
with the NMCSCF value (up to $4f$) of $-2.903117$ \cite{[13]}, the
exact value being $-2.903724$ \cite{[14]}.

2. For the $Li$ ground state $1s^{2}2s$ $^{2}S$, using
uncontracted correlation orbitals up to $4f'$, we obtain
$E_{15}^{u}$ $=$ $-7.4767$ $ a.u.$ \cite{[1]}, which is comparable
with the NMCSCF (up to $4f$) value of $-7.4762$ $a.u.$
\cite{[15]}, while the exact value is $-7.4780$ $a.u.$
\cite{[16]}.

3. For the $Li$ excited state $1s^{2}2p$ $^{2}P$, an example of
straightforward slow CI convergence, using contracted correlation
orbitals up to $4f$, we obtain $E_{13}^{c} =$ $ -7.4080$ $ a.u.$
\cite{[1]}, comparable with the large CI (45 CI terms up to $5g$)
value of $-7.4084$ $ a.u.$ \cite{[wwa63]}, while the experimental
value is $-7.4099$ $ a.u.$ \cite{[17]}.

4. For the $C$ ground state $1s^{2}2s^22p^2$ $^{3}P$, using
uncontracted correlation orbitals up to $4f$, including $1s'$,
$2s'$ and $2p'$ (13 orbitals), by keeping 64 mostly significant
configurations with 346 Slater determinants, we obtain
$E=-37.78719$ $ a.u.$, comparable with the value $E=-37.78695$ $
a.u.$ obtained by large-scale NMCHF  using up to $4f$ orbitals in
active space \cite{[sun91]}, and also with the value $E=-37.78885$
$ a.u.$ obtained by large-scale MRCI using 145 Gaussion functions
(17s11p6d5f4g2h) with $\thickapprox$ 1 500 000 configurations
\cite{[mav99]}. (With 90 orbitals and $\lessapprox$ 100 000 Slater
determinants Sundholm and Olsen obtain $E=-37.79$
a.u.\cite{[sun91]}, while Silverman \cite{[sil89]} has obtained
$E=-37.845$ $ a.u.$ using 1/Z expansions).

5. Finally, for He $1s2s$ $^{1}S$, by implementing the
Hylleraas-Undheim-MacDonald (HUM) theorem \cite{[11]} with these
orbitals, i.e. by optimizing the $2nd$ root of the secular
equation (the $1st$ would provide the ground state within the
\emph{same} basis functions), we obtained for $1s^{2}+2s$: $E_{2}
= -2.14261$ $a.u.;$ $1s^{2}+2s$ $(+1s')$: $E_{2} = -2.14389$
$a.u.$ and $1s^{2}+2s$ $(+1s')$ $+$ $2p$: $E_{2} = -2.14456 $
$a.u.$. The corresponding NMCSCF values are $1s^{2}+2s$: $E =
-2.14347$ $ a.u.$ and $1s^{2}+2s +2p$: $E = -2.14380$ $
a.u.$\cite{[13]}.

We observe that our values are quite close to NMCSCF, so that our
analytic orbitals and wavefunctions are quite trustable with
nearly as small CI expansions as NMCSCF. I.e. variationally moving
the nodes and the extent of the GLTOs they become quite similar to
NMCSCF orbitals with the same rich and concise physical content.

\section{Part II. An analysis of the general Eckart theorem}

The general Eckart \cite{[5]} theorem (GET) for excited states
(c.f. Appendix), states that: The exact energy eigenvalue $E_n$,
is a lower bound not of the calculated energy $E_e^{(n)}$ per se,
but of the calculated \emph{augmented} energy:
$(E_e^{(n)}+\delta_e^{(n)})$ $\equiv$ [$E_e^{(n)} + \sum\limits_{i
= 1}^{n-1} {|\langle\psi_i|\Psi_e^{(n)}\rangle|^2 (E_n-E_i)}$
$\geq E_n$]. Here $\psi _{1}$, $\psi _{2}$, ..., $\psi _{n}$, ...
are the exact eigenstates of the Hamiltonian $H$ with energies $
E_1 < E_2 < ... < E_n < ...$, and $\Psi _e^{(n)}$ is the
calculated normalized $(n-1)th$ excited state of the desired
symmetry, with energy expectation value $E_e^{(n)}$.

Even if the exact $\psi_i$ were used, the augmentation
$\delta_e^{(n)}$ \emph{would not be zero, for an approximate}
$\Psi_e^{(n)}$, so that, in trying to estimate the unknown $E_n$
via the minimization principle, by varying $\Psi_e^{(n)}$, the
augmentation $\delta_e^{(n)}$ should be taken into account.

Since $E_n = E_e^{(n)} + \delta_e^{(n)} - \epsilon_e^{(n)}$
[equation (\ref{En}) in Appendix], estimating $E_n$ requires
minimization of both $\delta_e^{(n)}$ and $\epsilon_e^{(n)}$
$\equiv$ $\sum\limits_{k=n+1}^{\infty} {|\langle\psi_{k}|\Psi
_e^{(n)}\rangle|^2 (E_k-E_n)}$, since both are unknown but
positive. This can be achieved if $\Psi_e^{(n)}\approx \psi_n$.
But then, since $\epsilon_e^{(n)}$ is positive and subtracted, we
can have a \emph{rather conservative} estimate of the error (the
energy uncertainty of $E_n$) by $\delta_e^{(n)}$ [which we
approximate by $\Delta_e^{(n)} \equiv \sum\limits_{i = 1}^{n-1}
{|\langle\Psi_e^{(i)}|\Psi_e^{(n)}\rangle|^2 (E_e^{(n)} -
E_e^{(i)})}$], provided that this dominates over the CI expansion
truncation (convergence) error $\varepsilon$. Evidently, for a
given $\Psi_e^{(n)}$, the closer all $\Psi_e^{(i)}$ are to the
lower lying $\psi_i$, the better is the estimation of
$\delta_e^{(n)}$ by $\Delta_e^{(n)}$. [In practice the subtraction
of $\epsilon_e^{(n)}$, which is dominated by a few closest to $n$
levels, and is, therefore, comparable to $\delta_e^{(n)}$, reduces
the error, making $E_e^{(n)}$, if $\Psi_e^{(n)}\approx \psi_n$,
much closer to the exact than the conservatively proposed
uncertainty $\Delta_e^{(n)}$]. Since variational collapse,
resulting to some large overlap
$|\langle\Psi_e^{(i)}|\Psi_e^{(n)}\rangle|^2$ for some $i < n$,
also reduces $\Delta_e^{(n)}$ via the $(E_e^{(n)} - E_e^{(i)})$
term, minimization of $E_e^{(n)}$ should be performed under both
restrictions that $|\langle\Psi_e^{(i)}|\Psi_e^{(n)}\rangle|^2$
(for all $i < n$) and $\Delta_e^{(n)} \equiv \sum\limits_{i =
1}^{n-1} {|\langle\Psi_e^{(i)}|\Psi_e^{(n)}\rangle|^2 (E_e^{(n)} -
E_e^{(i)})}$ be minimal.

In fact, since $\epsilon_e^{(n)}$ is never zero and always
unknown, an \emph{ab-initio} estimation of the uncertainty to
$E_n$ should always be attempted, even when minimizing by the HUM
theorem \cite{[11]}, which ensures $E_n \leq E_e^{(n)}$, for two
reasons: (i) Because some specific CI expansion might lead to
unacceptably large $E_e^{(n)}$. I.e. many trial functions must be
checked. (ii) Even if it happened that $E_e^{(n)} = E_n$,
$\Psi_e^{(n)}$ would be orthogonal to the lower lying roots,
\emph{the nth of them having been optimized}. Thus, the
orthogonality to the \emph{best} $\Psi_e^{(i)}$, $i<n$, i.e.
\emph{if the ith root had been optimized} for each $i$, is not
evident. That is, $\delta_e^{(n)}$ would not vanish, and it should
always be estimated.

\section{Part III. Ab-initio error estimation for excited states}

In the following we describe a (previously unreported) technique,
showing that it is possible, utilizing the ability for exact
orthogonality between GLTOs, to minimize the overlaps
$|\langle\Psi_e^{(i)}|\Psi_e^{(n)}\rangle|^2$ (and
$\Delta_e^{(n)}$), and thus obtain an \emph{ab-initio} estimate of
the energy uncertainty. Our technique is not a completely general
method for any (doubly, triply, etc) excited state, but at least
it is valid for singly excited states of (many electron) atoms; we
discuss below the limitations and disadvantages of our technique.
We demonstrate it in the simplest case of the 1st excited state
with 2 electrons and show its extension with some examples to
higher excited states and with more electrons.

\begin{figure*}
\includegraphics[width=5in]{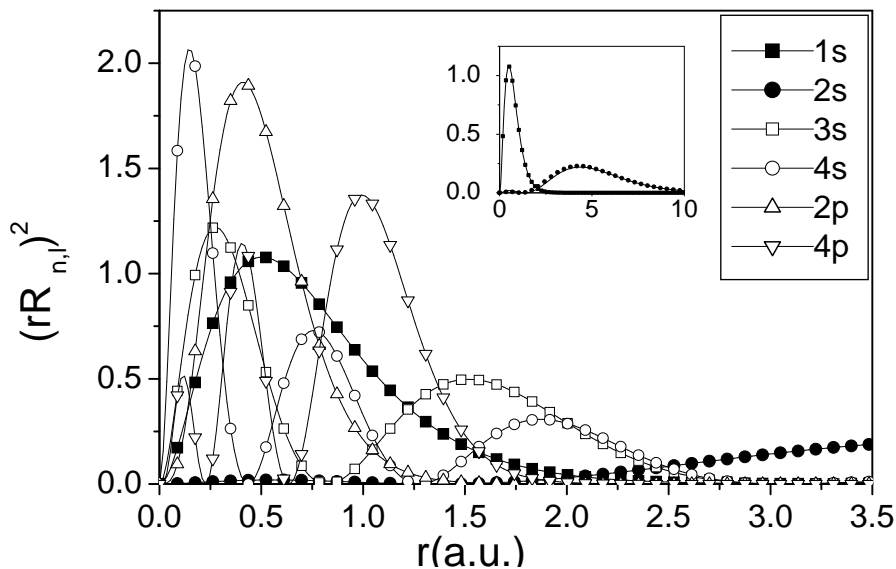}% Here is how to import EPS art
\caption{ Our full CI (up to $4f$) $He$ $1s2s$ $^{1}S$ orbitals.
The inset displays the `\emph{main}'  orbitals, both contracted
(lines), $E_{10}^{c} = -2.1459628$, $z_{1s} = 1.992$, $b_{1s} =
0.001$, $q_{1s} = 3.195$, $z_{2s} = 1.109$, $b_{2s} = 0.373$,
$q_{2s} = 1.683$, and uncontracted (symbols), $E_{10}^{u} =
-2.1459604$, $z_{1s} = 1.992$, $z_{2s} = 1.122$. We used the same
virtual orbitals for both cases; the most significantly
contributing have $z_{3s} = 4.372$, $z_{4s} = 8.257$, $z_{2p} =
9.775$, $z_{4p} = 24.483$.  All quantities are in $ a.u.$.
\label{f1} }
\end{figure*}

First a remark: Generally, if the CI expansion, instead of the
exact $\psi_n$, approximates a higher lying state, or collapses
quite lower than the exact $\psi_n$, these cases should be
rejected. The rejection may not be feasible with any variational
method, however, at least with the present GLTOs it is possible:
The \emph{higher} wave functions are easily recognizable by the
`\emph{main}' terms because GLTOs maintain their physical meaning
even after orthogonalization [c.f. figure (\ref{f1}) below] so
that they do not allow confusion with any fictitious simulation of
another (undesired) orbital \cite{[scffail_b],[scffail_m]} [for
example, whereas, with other methods 2s may be incorrectly
``approximated" by a 3s orbital, having two nodes and a very
shallow long tail, with the present method (via the GLTOs) this
cannot be confused with a correct 2s]. On the other hand, the
\emph{collapsed} wave functions have large overlap with some lower
$\Psi_e^{(i)}$ $\approx$ $\psi_i$ wave function.

In order to avoid variational collapse \cite{[collap]}, various
techniques exist in the literature. These invariably involve
either the state averaged NMCSCF approximation \cite{[sa]}, or
application of Hylleraas perturbation variation method
\cite{[pva]}, or the orthogonality constrained variation method
\cite{[ocva]}. We \emph{ab-initio} reject collapsed (and higher)
results by checking the various energy multiminima: Multiminima of
the energy surface can be visited by simulated annealing, which
has been used to determine the global energy minimum under
orthogonality constraints, without any effort to \emph{locate} the
various almost equivalent local energy minima \cite{[sia]}; in our
approach this is necessary, and is achieved by a thorough (or
guided, as explained below) search of \emph{the orbital parameter
space}.

So, let us consider first $\psi _{1}$, $\psi _{2}$, the exact
eigenstates of the ground and first excited states, with energies
$ E_1 < E_2$, and $\Psi_{g}$, $\Psi_{e}$ the corresponding
calculated approximations. Then
$  %\begin{eqnarray}
\Psi_{g} = \Psi_{g0} + h_{g}\phi_{g}, %\label{eqestim1}
$  %\end{eqnarray}
$  %\begin{eqnarray}
\Psi_{e} = \Psi_{e0} + h_{e}\phi_{e}, %\label{eqestim2}
$  %\end{eqnarray}
where $\Psi_{g0}$ and $\Psi_{e0}$ are `\emph{main}' wavefunctions,
(without correlation corrections) $h_{g}$ and $h_{e}$ are the
largest correlation coefficients, $\phi_{g}$ and $\phi_{e}$ are
corresponding calculated correlation corrections.

Supposing that we have achieved both: $\Psi_g$ be reliably close
to $\psi_1$ (comparable to NMCSCF) and $S_{eg} \equiv \left\langle
\Psi _e | \Psi _g \right\rangle $ be \emph{small} (as small as
possible), then $\Psi _e$ is almost orthogonal to the exact
$\psi_1$ with uncertainty $O(S_{eg})$. But it is not \emph{any}
state orthogonal to $\psi _1$: Since $E_2$ is discretely separated
from $E_1$, so there are no other energies in between, and higher
(and collapsed) functions are rejected, the (in terms of GLTO's
described) approximation (almost orthogonal to $\psi_1$) is close
to the exact $\Psi _e \approx \psi_2$ with enough accuracy, and
$E_2-E_1$ $\approx$ $E_e^{(2)}-E_g^{(1)}$. Then our estimated
(presumably minimal) $\Delta_e^{(2)}$ $\sim$ $O(S_{eg}^2(E_e^{(2)}
- E_g^{(1)}))$.

Furthermore, if our convergence criterion $\varepsilon \ll
\Delta_e^{(2)}$, then $E_2 \approx E_e^{(2)}$ +
$O(\Delta_e^{(2)})$, and if $\Delta_e^{(2)} \ll \varepsilon$, then
we have an even better approximation of $\Psi _e \approx \psi_2$,
and $E_2 \approx E_e^{(2)}$ + $O(\varepsilon)$.

As soon as we have determined $E_2$ and $\psi_2$ accurately enough
(by as small $S_{eg}$ and $\Delta_e^{(2)}$ as possible, i.e. by
the best possible $\Psi_e^{(2)}$, then we can consecutively
proceed to higher excited states (n), via all lower lying
\emph{best mutually (almost) orthogonal} calculated approximations
$\{\Psi_{e}^{(i)}$ ($i < n)\}$, \emph{provided that each $\psi_i$
is accurately enough resembled by $\Psi_{e}^{(i)}$}, determined as
above, by choosing the smallest overlaps $S_{ei}$ and
$\Delta_e^{(i)}$, consecutively for each i, until, for some $n$,
$\Delta_e^{(n)}$ becomes \emph{comparable} to the last energy
separation $E_n - E_{n-1}$. Depending on the accuracy of each
$\Psi_{e}^{(i)}$ and on the quality of the orthogonality of
$\Psi_{e}^{(n)}$ to each of them, due to error accumulation, at
about that $n$ this process becomes further unreliable.

In conclusion, in order to use orthogonalization to approximate
lower states, we need: (i) a trustable $\Psi_g \approx \psi_1$
($\Psi_e^{i} \approx \psi_i$), (ii) if possible, several well
converged $\Psi_{e}$'s (small $\varepsilon$'s) with: (iii) proper
well recognizable `\emph{main}' terms $\Psi_{e0}$'s and (iv)
minimal $S_{eg}$. Finally, we need to estimate not only the energy
by $E_e^{(n)}$ but also its augmentation by $\Delta_e^{(n)}$.
Prerequisites (i) and (ii) are evident; (iii) may not be possible
with any method, but with the present GLTOs it is automatically
fulfilled, after rejection of improper \emph{`main'} terms,
because these are recognizable and physically meaningful (as being
close to NMCSCF orbitals); (iv) is a problem:

%Supposing that, after appropriate rejections, we deal with proper
%and well converged wave functions,
We search for minimal $S_{eg}$ using our central idea to obtain
various good representations (close to NMCSCF) of the ground and
the excited states \emph{and to choose minimal overlap}. Rejecting
large overlaps $S_{eg}$ and improper `\emph{main}' terms will
exclude incorrect representations (higher or collapsed). Then from
all the accepted we should find the smallest $\Delta_e^{(n)}$.

Ideally the search for minimal $S_{eg}$ (and $\Delta_e^{(n)}$)
should require an exhaustive search for all (but, anyway, finite
in number) possible $S_{eg}$, which is out of our present
computational abilities. But since
\begin{eqnarray}
& S_{eg} =  \left\langle {{\Psi _{e0} }}|
 {{\Psi _{g0} }} \right\rangle + h_e \left\langle {{\Psi _{g0} }}|
 {{\phi _e }} \right\rangle +
 h_g \left\langle {{\Psi _{e0} }}| {{\phi _g }}
 \right\rangle \nonumber\\ & + O(h_eh_g),
\label{overlap_heg}
\end{eqnarray}
we prefer, if possible, to guide our search [and reduce it to a
linear process (than quadratic)] by observing that if we can
demand $\left\langle \Psi_{e0}| \Psi_{g0} \right\rangle = 0$, then
we need the smallest possible $h_e$ and $h_g$. This, with GLTOs,
can always be achieved for singly excited states if both
$\Psi_{g0}$ and $\Psi_{e0}$ can be described primarily by one or
more configurations from the same electron occupancy [or if it
happens that other contributing occupancies are already
(angularly) orthogonal]. Although this restricts the general
applicability of our (guided) method to such singly excited
states, it still covers many interesting cases, including
estimating uncertainties to typical occurrences of variational
collapse for which we shall give some examples.

We present first a demonstration of our idea in the $1s2s$ $^{1}S$
isoelectronic sequence from $He$ to $Ne$ and then we extend it to
some examples of higher singly excited states and with more
electrons.

Between 1936 and 1997, there have been many publications on $1s2s$
states, c.f. \cite{[13],[14],[18],[19],[20],[21]}. The most
accurate variational calculations, using Hylleraas \cite{[22]}
type trial functions, have been performed by the Pekeris group
\cite{[14]}. However, their method is of a non-central field type,
which cannot be straightforwardly extended to larger atoms, and
their wave functions, having more than 220 terms, do not provide a
simple understanding even for $He$. Fischer has performed NMCSCF
calculations of the $1s2s$ $^{1}S$ isoelectronic sequence from
$He$ to $Ne$ \cite{[4]}, with which we make comparisons in table
\ref{t1}.

By implementing the above guided search for $1s2s$ $^1S$ states,
$\Psi_{g0}\equiv 1s_{g}^{2}$ and $\Psi_{e0}\equiv 1s_{e}2s_{e}$,
the demand $ \left\langle 1s_g|2s_e \right\rangle = 0 $ makes
$\langle\Psi_{e0}|\Psi_{g0}\rangle = 0$. [In practice, we first
calculate $ \Psi _{g0} $ and obtain the optima $z_{1sg}$,
$b_{1sg}$, $q_{1sg}$, to replace the $z_{1s}$, $b_{1s}$, $q_{1s}$
of equation (\ref{g2s}), for every varied value of $z_{2se}$,
$b_{2se}$, $q_{2se}$]. \emph{If we use enough correlation
orbitals, then many slightly different $ \Psi _e $ ($ \Psi _g $),
having $\langle\Psi_{e0}|\Psi_{g0}\rangle = 0$, with almost the
same energy $E_e$ ($E_g$) (up to the $3rd$ decimal place) appear
as local energy minima } (not only the widest), in which the
coefficients $h_{e}$ ($h_{g}$) change slightly form minimum to
minimum. We use these minima to choose the smallest possible
$h_{e}$ and $h_{g}$, which makes $S_{eg}
% \equiv \left\langle {{\Psi _e }}| {{\Psi _g }} \right\rangle
$ as close to zero as possible [e.g. $10^{-2}$ a.u., and
$\Delta_e^{(n)}$ $~O(10^{-4})$].

Thus, for the ground state of $He$ $1s_{g} ^{2}$ $^{1}S$
correlated in full CI by $2s_{g}$, $3s_{g}$, $4s_{g}$, $2p_{g}$,
$3p_{g}$, $4p_{g}$, $3d_{g}$, $4d_{g}$, $4f_{g}$ orbitals (with
only one optimized $1s$ orbital in $\Psi_{g0}$ for exact
orthogonalization to $\Psi_{e}$), various significant candidate
minima were found (c.f. table \ref{cg}) from which we chose the
\emph{smallest} $h_{g }= 0.05$, ($z_{1sg}=1.6297$, $b_{1sg}=0.0$,
$q_{1sg}=1.0$; $z_{2sg}=4.6852$, $b_{2sg}=0.0$, $q_{2sg}=1.0$) for
orthogonalization to $\Psi_{e}$. Then we calculated the excited
state $1s_{e} 2s_{e}$ $^{1}S$ ($+3s_{e}$, $4s_{e}$, $2p_{e}$,
$3p_{e}$, $4p_{e}$, $3d_{e}$, $4d_{e}$, $4f_{e}$), and from the
most significant candidate minima (table \ref{cg}) we must choose
(reported in figure (\ref{f1})) the one with the \emph{smallest}
$h_{e} = 0.011$ (details are given in the caption) with $S_{eg} =
2.9\times10^{-2}$, so that our approximation to $\delta_{e}^{(2)}$
is $\Delta_{e}^{(2)}$ = $6.5\times10^{-4}$ a.u.. We also need
$\varepsilon$: The $s$, $p$, $d$, $f$ correlation for this
$\Psi_e^{(2)}$ (in a.u.) converges as $-2.14516$, $-2.14587$,
$-2.14595$, and $-2.14596$ respectively, (a faster converge than
by using HUM theorem); hence, since our $E_e$ value is well
converged by $\varepsilon \sim O(10^{-5})$, then the unknown $E_2
\approx E_e (= -2.14596$ a.u.) \emph{with uncertainty
$O(10^{-4})$}. I.e. given the accurate $\Psi_g \approx \psi_1$ and
all unreasonable representations to $\Psi_e$ having been excluded
(large $S_{eg}$ if collapsed, incorrect `\emph{main}' terms -
recognizable due to the GLTOs - if higher), the remaining being
\emph{necessarily} close to $\psi_2$, due to the discreteness of
the energy spectrum, via the closest of them: 4 digits of the
(unknown quantity) $E_2$ are guaranteed (without using external
information!) because the correction $\delta_{e}^{(2)}$ $\approx$
$\Delta_{e}^{(2)}$ [not regarding the subtraction of the positive
quantity $\epsilon_{e}^{(2)}$ (c.f. equation (\ref{En}) in
Appendix] starts after the 4th digit. In figure (\ref{f1}), the
virtual orbitals $2p$, $4p$ (and all others not displayed),
introduce by their lobes, an \emph{angular separation} between the
electrons, \emph{in places where the $1s$, $2s$ orbitals
appreciably overlap}, and the $3s$, $4s$ orbitals introduce nodes,
i.e., a \emph{radial separation} between the $1s$, $2s$ electrons,
all simulating the cups conditions. The good recognizability of
the main wavefunction, from the orbitals of figure (\ref{f1}) is
evident.

Similarly, we calculated the whole $1s2s$ $^{1}S$ isoelectronic
sequence from $He$ to $Ne$ (table \ref{t1}). We observe that our
values are quite comparable to NMCSCF (with seven configurations)
\cite{[4]} and approximate the exact \cite{[14]}. If the exact
were unknown, our \emph{ab-initio proximity estimates}
$\delta_e^{(2)}$ ($\approx$ $\Delta_e^{(2)}$) would guarantee at
least 3-4 decimal digits. The further ``coincidences" with the
exact energies, because of the nearly NMCSCF quality of GLTOs,
occur due to the subtraction of $\epsilon_e^{(n)}$ in (\ref{En})
(c.f Appendix). We could not find a better rigorous way to bracket
(locate) the unknown $E_n$ by taking more advantage of the high
quality of our orbitals other than minimizing first $S_{eg}$ and
then $\Delta_e^{(n)}$, because we could not estimate
$\epsilon_e^{(n)}$; however even with these uncertainty estimates,
bracketing the (unknown) exact $E_n$ to 3-4 guaranteed decimal
digits is enough to \emph{ab-initio} certify that the free $1s_{e}
\bot 2s_{e}$ values, shown in the last column of table \ref{t1},
are collapsed.

We also show an example of a higher excited state: He $1s3s$
$^1S$. By demanding $ \left\langle {{1s_g }}| {{3s_e }}
\right\rangle = 0 $ and $ \left\langle {{2s_{ep} }}| {{3s_e
}}\right\rangle = 0 $ (where $2s_{ep}$ is the previously
determined $2s_e$ orbital from the above 1s2s $^1S$ calculation),
we make $\Psi _{g0}$, $\Psi^{(2)} _ {(1s2s)0}$ and $\Psi^{(3)} _
{(1s3s)0}$ rigorously orthogonal to each other. Then, with the
same basis set as above, i.e., up to 4f orbitals, we obtain
$E_e^{(3)} = -2.06129$ a.u. with uncertainty $\Delta_e^{(3)}
\equiv \Delta_e^{(3,1)} + \Delta_e^{(3,2)} = 1.45 \times 10^{-4} +
2.85 \times 10^{-7} = 1.45 \times 10^{-4}$, which embraces the
(well known) exact value of -2.06127 a.u. \cite{[14]}. We should
mention that, although this value is by $O(10^{-5})$ below the
exact, it does not violate the GET, and the wave function might be
closer to the exact than another truncated approximation that
would approach the exact energy from above. However, if we free
all orbitals (without using the g-factors), we obtain -2.06859
a.u., which is out of our calculated uncertainty, therefore,
\emph{ab-initio} rejected as collapsed.

Finally, we give an example for more (three) electrons: Li
$1s(2s2p$ $^3P)$ $^2P$ [in the combination $1s\alpha$ $(2s\alpha
2p\beta$ + $2s\beta 2p\alpha)$ -  2 ($1s\beta 2s\alpha 2p\alpha$)
which is orthogonal to $1s(2s2p$ $^1P)$ $^2P$: $1s\alpha$
$(2s\alpha 2p\beta$ - $2s\beta 2p\alpha)$ ($\alpha,\beta$ mean
spin-up, spin-down)]. By demanding $\left\langle
2s_e|1s_g\right\rangle = 0 $, where $1s_g$ is the previously (test
3 of part I) determined $1s$ orbital from the lowest state of this
symmetry $1s^22p$ $^2P$, then, with the same basis set as above,
i.e., up to 4f orbitals, we obtain $E_e^{(2)} = -5.31998$ a.u.
with uncertainty $\Delta_e^{(2)}  = 7.54 \times 10^{-3}$, which
embraces the experimental value of -5.312 a.u. \cite{[rassi]},
while Weiss'es value in \cite{[ederer]} is -5.3111 a.u.. The
theoretical value of -5.331 a.u. (58.38 eV assigned (we think
incorrectly) to Goldsmith \cite{[golds]} in \cite{[rassi]}) is out
of our uncertainty estimate. If we free all orbitals (without
using the g-factors), we obtain -5.34139 a.u., which is out of our
calculated uncertainty, therefore, is also \emph{ab-initio}
rejected as collapsed.

\section{Conclusion}
In summary: (1) We presented previously unreported analytic GLTOs
which accurately and concisely describe (comparably with NMCSCF)
the correct atomic wave function. (2) We clarified the general
Eckart theorem for excited states concentrating on the importance
of the necessary augmentation $\delta_e^{n}$ to the calculated
energy. (3) Using this, we proposed a method to \emph{ab-initio}
bracket the (unknown) energy of singly excited states to some
significant digits. This gives some confidence as to where the
exact energy is located, and excludes collapsed outcomes,
\emph{without using external information}. Due to the accurate
description of the correct wave function of both the excited and
(even more important) the ground state, our method needs (i) the
good convergence of large CI expansions (ii) the potentiality
(feasible with analytic orbitals) for an exhaustive search in the
orbital parameter space, which is unavoidable in order to
determine minimal $S_{eg}$ and $\Delta_e^{(n)}$, and (iii) the
\emph{exact} mutual orthogonality of the excited $\Psi_e$ and
`\emph{main}' lower energy terms, leading directly to maximal
orthogonality of the total wave functions.

The present method, if guided by exact orthogonality of the
`\emph{main}' terms, is valid at least for singly excited states
in which all terms of $\Psi_0$ consist of Slater determinants of
the same occupancy. The less significant the `\emph{main}' terms,
the less useful (the ``guided" version of) the method.

For an \emph{ab-initio} estimate of the energy uncertainty
$\Delta_e^{(n)}$, an exhaustive search in orbital space seems
unavoidable for \emph{any} variational method able to provide a
clear orbital interpretation, like NMCSCF, by changing, e.g.,
starting values. We think that this should be tried by the
specialists: Since their wave functions are very accurate
($\Psi_e^{(i)}$ $\approx$ $\psi_i$), keep some small $S_{eg}$'s,
out of which the smallest $\Delta_e^{(n)}$ (some must be below
$E_n$), and estimate the unknown $E_n$ by $E_e^{(n)} +
\Delta_e^{(n)} \pm \varepsilon$ (!)

Our orbitals are being used in studying the radiative decay of
doubly excited states to singly excited states of He, where a
(previously unreported) good qualitative agreement for \emph{both}
the metastable atom and the $VUV$ photon spectra experiments
\cite{[oshm2000],[rc1999]} is obtained \cite{[23],[1]}.

\begin{acknowledgments}
Z. Xiong was partially supported by the Greek State Scholarship
Foundation (I.K.Y.) and by a National Hellenic Research Foundation
Scholarship.
\end{acknowledgments}

\newpage
\appendix*
\section{The General Eckart theorem}

Let $\psi _{1}$, $\psi _{2}$, ..., $\psi _{n}$, ... be the exact
eigenstates of the Hamiltonian $H$ (a complete orthonormal set)
with energies $ E_1 < E_2 < ... < E_n < ...$, and let
\begin{eqnarray*}
\Psi _e^{(n)} = \sum\limits_{i = 1}^{\infty} {\langle\psi
_{i}|\Psi _e^{(n)}\rangle \psi _{i} } ,
\end{eqnarray*}
with
\begin{eqnarray}
1 =\sum\limits_{i = 1}^{\infty} {|\langle\psi _{i}|
\Psi_e^{(n)}\rangle|^{2}} ,
 \label{anorm}
\end{eqnarray}
be the calculated normalized $(n-1)th$ excited state
approximation. Then
\begin{eqnarray}
E_e^{(n)} && =  \langle\Psi _e^{(n)}|H|\Psi _e^{(n)}\rangle \nonumber \\
&&  =  \sum\limits_{i = 1}^{n-1} {|\langle\psi _{i}|\Psi
_e^{(n)}\rangle|^2 E _{i} } + \sum\limits_{k = n}^{\infty}
{|\langle\psi _{k}|\Psi _e^{(n)}\rangle|^2 E _{k} }. \label{Een} \\
\nonumber
\end{eqnarray}
Multiplying (\ref{anorm}) by $E_n$ and subtracting form
(\ref{Een}) we obtain for (the unknown) $E_n$:
\begin{eqnarray}
E_n = E_e^{(n)} + \delta_e^{(n)} - \epsilon_e^{(n)}\label{En}
\end{eqnarray}
where both $\delta_e^{(n)}$ and $\epsilon_e^{(n)}$ are
\emph{positive} (or zero if $\Psi _e^{(n)}$ = $\psi _{n}$):
\begin{eqnarray}
\delta_e^{(n)} \equiv \sum\limits_{i = 1}^{n-1}
{|\langle\psi_{i}|\Psi_e^{(n)}\rangle|^2 (E_n-E_i)} \geq 0,
\label{del}\\
\epsilon_e^{(n)} \equiv \sum\limits_{k=n+1}^{\infty}
{|\langle\psi_{k}|\Psi _e^{(n)}\rangle|^2 (E_k-E_n)} \geq 0.
\label{ED}
\end{eqnarray}
Since it is impossible to calculate $\epsilon_e^{(n)}$, equations
(\ref{En}) and (\ref{ED}) imply that
\begin{eqnarray}
\epsilon_e^{(n)}=(E_e^{(n)}+\delta_e^{(n)}) - E_n \geq 0,
\label{get}
\end{eqnarray}
that is: \emph{The exact energy eigenvalue $E_n$, is a lower bound
of the calculated \emph{augmented} energy:}
$(E_e^{(n)}+\delta_e^{(n)})$ = [$E_e^{(n)} + \sum\limits_{i =
1}^{n-1} {|\langle\psi_i|\Psi_e^{(n)}\rangle|^2 (E_n-E_i)}$] - not
of just the calculated expectation value $E_e^{(n)}$. This is the
general Eckart theorem. For excited states the two terms
$\delta_e^{(n)}$ and $\epsilon_e^{(n)}$ in (\ref{En}) are
competing and $E_n$ may be either below or $above$ $E_e^{(n)}$,
unless $\Psi_e^{(n)}$ = $ \psi_n$ (which \emph{never} happens).
Therefore, \emph{any accidental equality, $E_e^{(n)}$ = $E_n$,
does not imply that $\Psi_e^{(n)} = \psi_n$, if $\delta_e^{(n)}$
$\neq 0$(!)} This should be kept in mind in any variational
calculation of excited states.

For the ground state, [$(n=1)$, i.e. e = g], (\ref{get}) reduces
to the usual Eckart upper bound theorem, since $\delta_g^{(1)}
\equiv 0$.

For excited states it means that between two approximate wave
functions lying slightly above and slightly below the exact
energy, the lower lying (with $E_e^{(n)} \lesssim E_n$) is more
trustable if it has less \emph{augmented energy} than the higher
lying(!). All lower lying approximations should not be generically
rejected; the one with the least \emph{augmented} energy is the
best approximation to $\psi_n$ (better than any higher lying).
(This seems not to have been adequately realized in the
literature).

%\newpage
\begin{table}
\caption[]{

An example of \emph{CI convergence} using the proposed GLTOs: The
CI convergence of $He$ $1s^{2}$, compared with NMCSCF, and
characteristics of our converged orbitals. The RMS extent is
analytically related to $z_{nl}$. The $nl'$ orbitals are
\emph{non-orthogonal} to the others ($nl$). Atomic units are
used.} \label{tab:orb}
\begin{ruledtabular}
\begin{tabular}{cccccccc}
  $nl$ & $nl'$ & $z_{nl}$ & $z_{nl'}$ & $RMS$ & $RMS'$ & E & NMCSCF \\
 \colrule
    1s  & 1s' &  1.4193   & 2.5517  & 1.2204 & 0.6788 & -2.87689  & -2.86168 \\
    2s  &     &  7.7657   &         & 0.4648 &        & &  \\
    2p  & 2p' &   8.9698  & 8.4431  & 0.6106 & 0.6487 & -2.89978  &  -2.89767 \\
    3s  &     &  12.1491  &         & 0.5218 &        & &  \\
    3p  &     &   9.2664  &         & 1.4450 &        & &  \\
    3d  & 3d' &  13.3661  & 11.9619 & 0.8398 & 0.9384 &  -2.90242  &  -2.90184 \\
    4s  &     &  18.5495  &         & 0.5114 &        & &  \\
    4p  &     &  19.5524  &         & 0.6490 &        & &  \\
    4d  &     &  21.2589  &         & 0.8563 &        & &  \\
    4f  & 4f' &  18.3067  & 26.0011 & 1.0364 & 0.7297 &  -2.90310  &  -2.90291 \\
%\hline
\end{tabular}
\end{ruledtabular}
\end{table}

\begin{table}
\caption[]{

The entries of this table are used as a tool to finally estimate
an uncertainty to the exact energy for the excited states: Various
significant candidates for the He ground state $ \Psi_g =
1s_{g}^{2}$ $^{1}S$ and for the first excited state $ \Psi_e =
1s_{e} 2s_{e}$ $^{1}S$, correlated in full CI up to $4f_{g}$ and
$4f_{e}$ respectively. We report in figure (\ref{f1}) the $\Psi_e$
with the smallest $h_{e} = 0.011$, orthogonalized to the $\Psi_g$
with the lowest $h_{g} = 0.05$, corresponding to the smallest $
\langle \Psi _e | \Psi _g \rangle = 2.9 \times 10^{-2}$. }
 \label{cg}
\begin{ruledtabular}
\begin{tabular}{lcc}
Main configurations of $\Psi_g$ & $z_{1sg}$ & $E_{g} = E_{e}^{(1)}$ \\
  \colrule
  $0.98[1s_{g} ^{2}] - 0.17[1s_{g}2s_{g}] + ... $ & 1.4080 & -2.9028\\
  $0.98[1s_{g} ^{2}] - 0.16[1s_{g}2s_{g}] + ... $ & 1.4057 & -2.9014\\
  $0.99[1s_{g} ^{2}] - 0.05[1s_{g}2s_{g}] + ... $ & 1.6297 & -2.9015\\
%\hline
 \colrule
 Main configurations of $\Psi_e$&& $E_{e}^{(2)}$\\
 \colrule
 $0.999 [1s_{e} 2s_{e}] + 0.017 [2p_{e}3p_{e}] + ... $ &&  -2.14604\\
 $0.999 [1s_{e} 2s_{e}] + 0.011 [3s_{e}^{2}] + ... $   &&  -2.14596\\
 $0.936 [1s_{e} 2s_{e}] + 0.350 [1s_{e} 3s_{e}] +... $ &&  -2.14583\\
\end{tabular}
\end{ruledtabular}
\end{table}

\begin{table*}
\caption[]{

Our full $CI$ up to $4f$ energies for the $1s2s$ $^{1}S$
isoelectronic sequence from $He$ to $Ne$, and of other excited
states, compared with other calculations (in a.u.). We used 10
uncontracted orbitals. In the \emph{ab-initio} proximity
estimation to $E_2$, we approximate $\delta_e^{(2)}$ by
$\Delta_e^{(2)}$. $6.5^{-4}$ means $6.5 \times 10^{-4}$. The free
one-configuration $1s2s$ values (i.e. $1s_{e} \bot 2s_{e}$), in
the last column \emph{should be ab-initio rejected} by our method,
as collapsed.}
 \label{t1}
\begin{ruledtabular}
\begin{tabular}{lrrrrlr}
 & $ E_e^{(2)}$ & Exact\cite{[14]}
 & MCSCF\footnotemark[1] & $z_{1sg}$
 & $\Delta_e^{(2)}$& $1s \bot 2s$\\
 \colrule
He I& -2.14596\footnotemark[2]& -2.14597&
-2.14595\footnotemark[3]&
1.6297& $6.5^{-4}$&  -2.156 \\
Li II& -5.04093& -5.04087& -5.04028& 2.4353 & $1.7^{-3}$&  -5.058\\
Be III& -9.18469& -9.18487& -9.18413& 3.6849 &$ 2.0^{-4}$&  -9.206 \\
B IV& -14.57834& -14.57853& -14.57769& 4.4797 &$ 3.7^{-3}$& -14.603 \\
C V& -21.22258& -21.22202& -21.22111& 5.5464 &$ 1.2^{-3}$& -21.248   \\
N VI& -29.11382& -29.11542& -29.11445& 6.4736 & $1.7^{-3}$& -29.143   \\
O VII& -38.25841& -38.25876& -38.25775& 7.4671 & $1.9^{-3}$& -38.288  \\
F VII& -48.65206& -48.65206& -48.65102& 8.4689 &$1.9^{-3}$& -48.682  \\
Ne IX& -60.29534& -60.29534& -60.29428& 9.4527 &$ 2.7^{-3}$& -60.327 \\
\hline He 1s3s $^1S$ & -2.06129 & -2.06127 & -2.06127
\cite{[chen94]} & 1.1090\footnotemark[4] &
$1.5^{-4}$$^{(}$\footnotemark[5]$^{)}$
& -2.069\footnotemark[6]\\
\hline Li $1s(2s2p$ $^3P)$ $^2P$ &  -5.31998 & -5.312
\cite{[rassi]}
& -5.3111\footnotemark[7] & 2.9670 & $7.54^{-3}$ &  -5.341\footnotemark[6]\\
%$1s(2s2p$ $^3P)$ & & & & & & \\
\end{tabular}
\end{ruledtabular}
%\footnotetext[1]{Accad et al. Reference \cite{[14]}.}
\footnotetext[1] {Froese Fischer C., Reference \cite{[4]} (with
seven configurations).}
\footnotetext[2] {c.f. The caption of figure (\ref{f1}).}
\footnotetext[3] {Froese Fischer C. et. al,
Reference \cite{[13]}, p. 67, (up to 6h).}
%\footnotetext[5] {Chen M-K, Reference \cite{[chen94]}. }
\footnotetext[4] {The z of $2s_{ep}$; that of $1s_{g}$ is shown in
the first line of He I.}
\footnotetext[5] {The uncertainty
$\Delta_e^{(3)} \equiv \Delta_e^{(3,1)} + \Delta_e^{(3,2)}$.}
\footnotetext[6] {Free variation (without $g_k$-factors).}
\footnotetext[7] {Weiss in \cite{[ederer]}}
\end{table*}

\begin{references}
\bibitem{[10]} We modified Powell's method
p. 412 of ref. \cite{[6]} for a restricted region in parameter
space.
\end{references}
\end{document}